\documentclass[conference]{IEEEtran}
\usepackage{amsmath}
\usepackage{amssymb}
\usepackage{tabularx,booktabs}
\usepackage{array}

\ifCLASSINFOpdf
  \usepackage[pdftex]{graphicx}
  \DeclareGraphicsExtensions{.pdf,.jpeg,.png}
\else
  \usepackage[dvips]{graphicx}
  \graphicspath{{../eps/}}
  \DeclareGraphicsExtensions{.eps}
\fi
\ifCLASSOPTIONcompsoc
  \usepackage[caption=false,font=normalsize,labelfont=sf,textfont=sf]{subfig}
\else
  \usepackage[caption=false,font=footnotesize]{subfig}
\fi
\begin{document}
\title{Data-Rate-Aware High-Speed CNN Inference on FPGAs}

%\author{\IEEEauthorblockN{Blinded for review}
%\IEEEauthorblockA{
%}

\author{\IEEEauthorblockN{Tobias Habermann}
\IEEEauthorblockA{Department of Applied Computer Science
\\Fulda University of Applied Sciences \\
Fulda, 36039 \\
Germany\\
Email: tobias.habermann@informatik.hs-fulda.de}
\and
\IEEEauthorblockN{Martin Kumm}
\IEEEauthorblockA{Department of Applied Computer Science
\\Fulda University of Applied Sciences \\
Fulda, 36039 \\
Germany\\
Email: martin.kumm@informatik.hs-fulda.de}
}


%Telephone: (800) 555--1212, Fax: (888) 555--1212}


\newcolumntype{C}[1]{>{\centering\arraybackslash}p{#1}} % centering column type with fixed width
\newcolumntype{R}[1]{>{\raggedleft\arraybackslash}p{#1}} % right aligned column type with fixed width
\newcolumntype{L}[1]{>{\raggedright\arraybackslash}p{#1}} % left aligned column type with fixed width

\maketitle

\begin{abstract}
Dataflow-based CNN accelerators on FPGAs achieve low latency and high throughput by mapping computations of each layer directly to corresponding hardware units. However, layers such as pooling and strided convolutions reduce the data at their output with respect to their input, strongly effecting the data rate of the following layers. This leads to underutilization in fully unrolled designs. While prior work introduced data-rate-aware layer-wise adaptation, determining the most efficient implementation remains challenging.

This paper presents a data-rate-aware CNN accelerator architecture for multi-pixel processing. Building on existing analytical models, the proposed method performs design-space exploration to identify configurations that improve hardware utilization and resource efficiency while preserving continuous flow of data, keeping all hardware units busy. Experimental results show substantial reductions in arithmetic resources compared to previous designs, enabling efficient implementation of complex CNNs on a single FPGA across a wide range of data rates.
\end{abstract}


% \fi
\IEEEpeerreviewmaketitle

\section{Introduction}

Convolutional neural networks (CNNs) are widely used in applications requiring low-latency and high-throughput inference, such as autonomous driving~\cite{selfdriving_challanges}, speech recognition~\cite{speechreq}, and high energy physics~\cite{engels_work}. 
Field-programmable gate arrays (FPGAs) are well suited for these workloads due to their ability to exploit fine-grained parallelism and application-specific customization. 

FPGA-based accelerators can be broadly divided into three
architectural classes. 
There are fully parallel designs, where the entire input image is processed in parallel~ \cite{umuroglu2020logicnets,andronic2023polylut,andronic2024neuralut,lou2024polylut,wang2019lutnet}. 
They achieve very low latency, but do not scale to larger networks and cannot implement convolutional layers. 
Then, dataflow architectures like FINN~\cite{umuroglu2017finn} or hls4ml~\cite{hls4ml} which process the image sequentially which improves scaling and allow for bigger CNNs to be implemented, but they are not data-rate aware and have a slower target speed.
Lastly there are high-speed, data-rate aware architectures like \cite{main_ref, old_main_ref}, where each layer implementation is tailored to the specific input data rate of that layer.
This data-rate-aware layer implementation is important when implementing CNNs as layers such as pooling and strided convolutions reduce the data rate along the processing pipeline, leading to underutilization in fully parallel designs.
%Among FPGA-based inference accelerators, dataflow and unrolled architectures provide high throughput by mapping neural network operations directly to hardware. 

Prior work introduced a data-rate-aware, continuous-flow CNN design paradigm~\cite{main_ref} that adapts the parallelization and resource sharing of individual layers to the local data rate, enabling high hardware utilization. 
While this approach achieves a data-rate-matched layer implementation by analyzing the data flow in the model, the approach is not designed to process more than one pixel per clock cycle. %it does not explore the full range of viable data-rate-matched layer implementations that can be realized when considering the data rate.

This paper extends the continuous-flow paradigm by adapting the approach for multi-pixel processing and simplifies the paradigm by condensing the implementation parameters and constrains.
%Rather than targeting a single input data rate, the proposed method explores all feasible data rates that fully utilize the first layer and iteratively refines layer implementations throughout the network. 
%By combining analytical data-rate models with design-space exploration, the framework identifies continuous-flow CNN architectures that improve hardware utilization and resource efficiency across a wide range of data rates. 

 

\newcommand{\rout}{r_{\ell}}
\newcommand{\rin}{r_{\ell-1}}
\newcommand{\dout}{d_{\ell}}
\newcommand{\din}{d_{\ell-1}}
\newcommand{\sigout}{s_{\ell}}
\newcommand{\sigin}{s_{\ell-1}}

\newcommand{\Crin}{C_{r_{\text{in}}}}
\newcommand{\Crout}{C_{r_{\text{out}}}}

\newcommand{\jmax}{j_\text{max}}
\newcommand{\hmax}{h_\text{max}}
\newcommand{\hbest}{h_\ell}
\newcommand{\jbest}{j_\ell}
\newcommand{\allH}{H_\ell}
\newcommand{\allJ}{J_\ell}
\newcommand{\floor}[1]{\left\lfloor #1 \right\rfloor}
\newcommand{\ceil}[1]{\left\lceil #1 \right\rceil}
\newcommand{\setof}[1]{\left \{ #1 \right \}}
\newcommand{\minof}[1]{\text{min} \left ( #1 \right )}

%\subsection{Background}

%Describe the general background here ...

%\begin{equation}
%\label{eq.rout} 
%\rout = \frac{\dout \cdot \rin}{\din \cdot s^2}
%\end{equation}
%Describe the previous work, so, the idea and inner workings of the journal paper... 
%Say that the previous work showed that some model architectures can lead to underutilization, as layer implementation settings like the number of configurations or data interleaving have to be rounded up to ensure a valid design.

\section{Adapting Continuous-Flow} % This subsection name is bad :D

\subsection{Continuous-Flow architecture}
This work builds upon \cite{main_ref}, which is described in the following. Each layer $\ell$, with $\din$ input channels and $\dout$ output channels (aka filters or neurons), in the model is implemented in sequence.

\begin{figure}[t]
\centering
\includegraphics{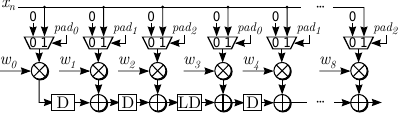}
\caption{The KPU base component presented in \cite{main_ref}.}
\label{fig.KPU_basic}
\end{figure}

Convolutional, and depthwise convolutional layers are implemented using a base component called kernel processing unit (KPU). The KPU takes the kernel weights and computes one sliding window of a feature map per clock cycle.
An example of a KPU that processes a $3 \times 3$ kernel is shown in Fig.~\ref{fig.KPU_basic}
where each input is multiplied with the kernel weights and then summed up with previous weighted features.
Each column of multipliers share the same padding select signal $pad_i$ which allows implicit zero padding at the sides of the sliding window.
By reconfiguring the weights and adjusting the delays, the KPU can process multiple kernels in sequence.

\begin{figure}[t]
\centering
\includegraphics{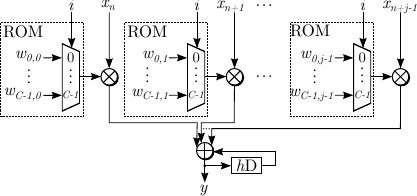}
\caption{The FCU base component presented in \cite{main_ref}.}
\label{fig.FCU_basic}
\end{figure}

The pointwise convolutional, and fully connected layers are implemented using a base component called the fully connected unit (FCU) which processes multiple neurons in sequence.
An example is shown in Fig.~\ref{fig.FCU_basic}
where $j$ features are processed per clock cycle and $h$ neurons are processed sequentially.
The input is held for $h$ clock cycles until the next $j$ features are processed, until all $\din$ features are processed.

Based on the input data rate into the layer $\rin$, which describes how many features have to be processed per clock cycle, a data-rate-matched layer implementation is derived.

For convolutional layers, this is achieved by computing the number of reconfigurations $C$ of the arithmetic-units and the data interleaving factor $I$ for the current layer. 
The number of configurations $C$ is directly derived from $\rin$ by
\begin{equation}
\label{eq.kpu_configs_amount}
C = \min \left (\left \lceil \frac{\din}{\rin} \right \rceil, \din \dout \right)
\end{equation}
and the interleaving factor $I$ is derived from $C$ by
\begin{equation}
\label{eq.interleaving_amount}
I = \left \lceil \frac{C}{\din} \right \rceil \, .
\end{equation}
In some scenarios, this leads to rounding errors which results in an architecture that is underutilized.

For fully connected and pointwise convolutional layers, the input data rate $\rin$ is first split up into its numerator $\jmax$ and denominator $\hmax$. The layer implementation is then defined to process $\jbest$ inputs over  
\begin{equation}
\label{eq.best_fcu}
\hbest = \text{max}(\{ h \in \mathbb{N} \, | \, h \text{ divides } \dout, h \leq \hmax \}) 
\end{equation}
clock cycles.

\subsection{Improved Continuous-Flow Architecture}

In the following, we will condense the approach shown in \cite{main_ref} for the data-rate-matched layer implementation starting with fully connected and pointwise convolutional layers and then generalize the approach for convolutional layers and other layer types.
Then, we will define the constraints of the architecture parameters, and extend the architecture to process multiple pixels at once.
% This allows a better resource allocation and allows more data rates

The Continuous-Flow architecture presented in \cite{main_ref} for the fully connected and pointwise convolutional layers can be condensed to the parameters $j$ and $h$. 
\begin{figure}[t]
  \centering
  \subfloat[Fully connected]{
    \includegraphics[width=0.37\linewidth]{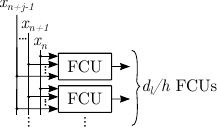}
    \label{fig.FC_basic}
  }
  %\hspace{0.01\linewidth}   % maximize separation between the subfigures
  \subfloat[Convolutional]{
    \includegraphics[width=0.52\linewidth]{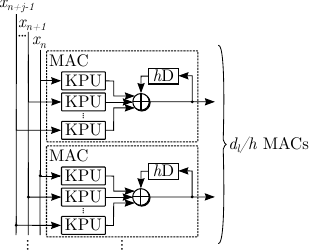}
    \label{fig.Conv_basic}
  }
\caption{The structure of convolutional and fully connected layers.} 
\label{fig.both_basic}
\end{figure}

%\begin{figure}[t]
%\centering
%\includegraphics[width=5cm]{figures/FC_basic.pdf}
%\caption{The condensed structure of fully connected and pointwise convolutional layers in \cite{main_ref}.}
%\label{fig.FC_basic}
%\end{figure}
Fig.~\ref{fig.FC_basic} shows the layer architecture for fully connected and pointwise convolutional layers. Each FCU processes $h$ neurons, so there are $\dout/h$ FCUs in the layer where all FCUs share the same $j$ input signals. 
%Each FCU calculates the current weighted input for the current neuron it is processing. 
%All $j$ input signals are held for $h$ clock cycles, so the FCUs can process the input for each neuron sequentially. 

%The output of a neuron is determined by the weighted sum of all $\din$ features of the current pixel.
In total, each FCU has to cycle through 
\begin{equation}
\label{eq.C} 
C = \frac{h \cdot \din}{j}
\end{equation}
configurations of weights to calculate each output of its $h$ neurons.
So with $C$ configurations, the layer produces
\begin{equation}
\label{eq.Crout} 
\Crout = \frac{h \cdot \frac{\dout}{h}}{C}=\frac{\dout \cdot j}{\din \cdot h}
\end{equation}
outputs per clock cycle and processes
\begin{equation}
\label{eq.Crin} 
\Crin = \frac{j}{h}
\end{equation}
inputs per clock cycle.

The same approach can be applied on the convolutional layer. 
%\begin{figure}[t]
%\centering
%\includegraphics[width=5cm]{figures/Conv_basic.pdf}
%\caption{The condensed structure of the convolutional and depthwise convolutional  layers in \cite{main_ref}.}
%\label{fig.Conv_basic}
%\end{figure}
Fig.~\ref{fig.Conv_basic} shows how to describe the convolutional layer with $j$ input signals, where each KPU computes $h$ different kernels. 
The KPUs are grouped together into multiply accumulate (MAC) units, where all outputs are accumulated, and partial results are stored. 
Each MAC unit consists of $j$ KPUs and is very similar to the FCU itself.
By removing the adders in the MAC units, the most common depthwise convolution with a group size equal to $\din$ can be described too, where the channel multiplier replaces $\dout$ when calculating the number of MACs and the upper limit for $h$. 
Other layer types, for example pooling layers, can be described by adapting this depthwise convolution setup further by replacing the KPUs with the base component for pooling layers described in \cite{main_ref}.

\subsection{Defining the Constrains of $j$ and $h$}
In \cite{main_ref} the layer parameters for the $\ell$-th layer, $\jbest$ and $\hbest$, were calculated based on $\jmax$ and $\hmax$ where $\rin=\frac{\jmax}{\hmax}$. 
We propose to reformulate this problem similar to an upper diophantine approximation of $\rin$ with constrained nominator $\jbest$ and denominator $\hbest$. 
To ensure that the FCUs do not have to process padded or invalid data, the number of input signals into the layer $j$ should be divisible by $\din$
\begin{equation}
\label{eq.bigJ} 
\allJ = \setof{ j \in \mathbb{N} \, | \, j  \text{ divides } \din } \, .
\end{equation}
Vice versa, the number of neurons processed in each FCU $h$ should be divisible by $\dout$ to ensure that all FCUs process the same number of neurons to eliminate empty times due to synchronization between FCUs, i. e.,
\begin{equation}
\label{eq.bigH} 
\allH = \setof{ h \in \mathbb{N} \, | \, h \text{ divides } \dout } \, .
\end{equation}
In total, these constrains ensure that if the layer is provided with enough data, the arithmetic units will always process valid data without any empty times wasted for synchronization or additional control circuitry to insert padding data, or filter out invalid data.

\subsection{Defining the layer implementation parameters of layer $\ell$}
All viable parameter settings for the layer $\ell$ that satisfy $\rin$ are then described by 
\begin{equation}
\label{eq.bigHJ} 
\text{HJ}_l = \setof{ (j,h) \, | \, j \in J \wedge h \in H \wedge \frac{j}{h} \geq \rin } \, .
\end{equation}
The most important criteria for choosing the right implementation setting is how close the implementation gets to the actual input data rate $\rin$ which is described by
\begin{equation}
\label{eq.BestRate} 
\text{BestRate}(\text{HJ}_l) = \minof{ \setof{ \frac{j}{h} \, | \, (j, h) \in \text{HJ}_l }} \, .
\end{equation}
Finally, a parameter setting in $\text{HJ}_l$ is selected where $\frac{j}{h}$ is closest to $\rin$
\begin{equation}
\label{eq.BestRateSelect} 
(\jbest, \hbest) \in \setof{ (j,h) \, | \, (j, h) \in \text{HJ}_l \wedge \frac{j}{h} = \text{BestRate}(\text{HJ}_l) } \, .
\end{equation}
In general, it is best to select a parameter setting where $h$ is close do $\dout$ to reduce the number of components (for example KPUs or FCUs) in the layer. This leads to just a few components with big adder trees which can be combined into resource efficient compressor trees~\cite{8263391}.
%It is important to note that if DSPs are used for multiplication, keeping the number of components to at least two rather than one. Otherwise the DSPs cannot ... bla bla dual Mult dsps and stuff...
%Note that the input aggregation is not needed as presented in \cite{main_ref}. 
These constrains allow for a better adaptation of the layer to the input data rate rather than in \cite{main_ref} where rounding could occur and the input aggregation is constrained. 
%For example, in the presented mobilenetV1 in \cite{main_ref}, the last layer has an input data rate of $\frac{1}{49}$ ...

\subsection{Adapting the Architecture for Multi-pixel Processing}
%The work in \cite{main_ref} is extended towards multi-pixel processing in the following. 
%\begin{figure}[t]
%\centering
%\includegraphics[width=5cm]{figures/FC_mp.pdf}
%\caption{An example of a pointwise convolutional or fully connected layer that can process two pixels per clock cycle.}
%\label{fig.FC_mp}
%\end{figure}
%Fig.~\ref{fig.FC_mp} shows how the pointwise convolution and fully connected layers are adapted to process two pixels at once. 
%The term that represents the current input, $x_n$, has to be adapted to $x_{n,m}$ to address the current pixel index.
%As the figure shows, only the number of FCUs doubles, where each FCU is processing a specific pixel.
To allow pointwise convolution and fully connected layer implementations to process two pixels at once, the number of FCUs doubles, where each FCU is processing a specific pixel.

Adapting the convolutional and depthwise convolutional layer to process multiple pixels is more challenging. 
\begin{figure}[t]
\centering
\includegraphics[width=5.5cm]{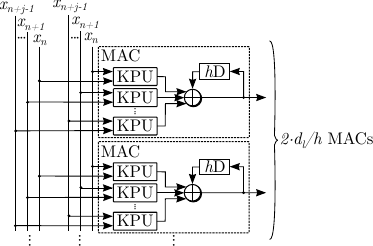}
\caption{An example of a convolutional layer implementation that can process two pixels per clock cycle.}
\label{fig.Conv_mp}
\end{figure}
Fig~\ref{fig.Conv_mp} shows an example of a convolutional layer where two pixels are processed per clock cycle.
The term that represents the current input, $x_n$, has to be adapted to $x_{n,m}$ to address the current pixel index too.
Each KPU has to process both pixels to process the current sliding window for the current kernel. 
Thereby, the KPU has to be adapted to be able to process two pixels at once.

\begin{figure}[t]
\centering
\includegraphics{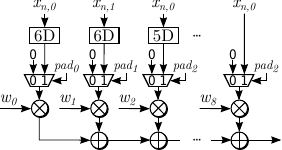}
\caption{A non-transposed KPU that can process two pixels per clock cycle.}
\label{fig.KPU_mp}
\end{figure}

An example of an adapted KPU for a $5 \times 5$ feature map with a $3 \times 3$ kernel is shown in Fig.~\ref{fig.KPU_mp}.
The adapted KPU is a non-transposed version of the KPU presented in \cite{main_ref}. 
So, rather than buffering the weighted partial results, the non-transposed KPU buffers the input features which can be buffered once, and then shared with all other KPUs in the layer.
Also, the adders at the bottom shown in Fig.~\ref{fig.KPU_mp} can be reduced into a compressor tree~\cite{8263391}.
The input data into the multipliers is delayed to make sure that all multipliers calculate all inputs of the current sliding window at the same time.
\begin{table}[t]
\centering
\caption{The resources used to implement the MobileNetV1 model with the same data rate. }
\label{td.compare_to_mainref}
\begin{tabular}{cccccc}
\toprule
\textbf{name} & \textbf{LUT} & \textbf{FF} & \textbf{BRAM} & \textbf{URAM} & \textbf{DSP}   \\
\cmidrule(lr){0-0} \cmidrule(lr){2-6}
\cite{main_ref}        & 204,931    & \textbf{563,255}    & 1,702.5    & \textbf{0}          & 5,691      \\
%\cite{main_ref}       & 215,465    & 726,972    & 1,454.5    & 10         & 5,691      \\
Ours                 & \textbf{158,540}    & 603,372    & \textbf{1,449.5}    & 10         & \textbf{5,664}      \\
\bottomrule
\end{tabular}
%\\
%\vspace{1mm}
%\footnotesize{*This implementation was created using the newest Version of our code generator.}
\end{table}
\begin{table*}[t]
\centering
\caption{The same MobileNetV2 CNN model implemented for different data rates compared to the SOTA.}
\label{td.mobnetv2_datarates}
\begin{tabular}{cccccccccccc}
\toprule
%\textbf{name} & \textbf{$\text{F}_{\text{max}}$} & \textbf{FPS} & \textbf{Latency} & \textbf{LUT} & \textbf{FF} & \textbf{BRAM} & \textbf{URAM} & \textbf{DSP}   \\
\textbf{} & \textbf{$\text{F}_{\text{max}}$} & \textbf{FPS} & \textbf{Latency} & \textbf{LUT} & \textbf{BRAM} & \textbf{URAM} & \textbf{DSP} & \textbf{Power} & \textbf{P. Eff.} & \textbf{Precision} & \textbf{FPGA} \\
      & MHz & Inf/s & ms &  &  & &  & W & mJ/Inf. &  & \\
 \cmidrule(lr){2-4} \cmidrule(lr){5-8} \cmidrule(lr){9-10}
%FINN \cite{finn_r}         & FREQ       & FPS        & Latency    & LUTs       & BRAM      &URAM& DSP    & Power & P/Inf  & Prec  & FPGA        \\
RTX 3080 GPU**              & 1860       & 4105.4     & 14.93      & -          & -         & -  & -      & 328   & 79.89  & 32-bit&         \\
LUTMUL \cite{lutmul_ref}    & 333        & 1627       & -          & 529k       & 1119      & -  & 106    & 42.12 & 25.89  & 4-bit & Alveo U280  \\
\cite{old_main_ref}         & 241        & 4803.1     & 0.62       & 136k       & 1225.5    & -  & 2709   & 18.04 & 3.75   & 8-bit & XCVU37P     \\
FINN \cite{finn_r}*         & 333        & 925        & -          & 501k       & 898       & -  & 106    & 41.69 & 45.07  & 4-bit & Alveo U280  \\
Ours (6/1)           & 403.71     & 16,020.40  & 0.21       & 186k       & 1,410      & 12         & 6,302      & 92.34      & 5.76       & 8-bit      & XCVU37P    \\
Ours (3/1)           & 404.53     & 8,026.40   & 0.42       & 124k       & 1,194.5    & 4          & 3,168      & 57.01      & 7.10       & 8-bit      & XCVU37P    \\
Ours (3/2)           & 400.64     & 3,974.61   & 0.85       & 77k        & 1,038      & 30         & 1,765      & 35.62      & 8.96       & 8-bit      & XCVU37P    \\
Ours (3/4)           & 405.52     & 2,011.48   & 1.66       & 52k        & 1,048      & 19         & 928        & 24.87      & 12.36      & 8-bit      & XCVU37P    \\
Ours (3/8)           & 408.33     & 1,012.72   & 3.30       & 41k        & 1,063.5    & 25         & 526        & 19         & 18.76      & 8-bit      & XCVU37P    \\
Ours (3/16)          & 410.00     & 508.44     & 7.54       & 33k        & 1,068      & 26         & 306        & 16.93      & 33.30      & 8-bit      & XCVU37P    \\
Ours (3/32)          & 353.48     & 219.17     & 14.92      & 30k        & 1,078      & 21         & 212        & 14.56      & 66.43      & 8-bit      & XCVU37P    \\
\bottomrule
\end{tabular}
    \\
    \vspace{1mm}
    \footnotesize{*FINN implementation results are from \textbf{\cite{lutmul_ref}}}
    \footnotesize{**Batch size was set to 832, only for the latency the batch size was set to 1}
\end{table*}
By analyzing the input image, it can be determined which input signal should be connected to a multiplier and how long the input has to be delayed.
\begin{figure}[h]
  \centering
  \subfloat[]{
    \includegraphics[width=0.22\linewidth]{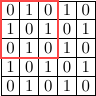}
    \label{fig.KPU_mp_data_order_a}
  }
  \hspace{0.1\linewidth}   % maximize separation between the subfigures
  \subfloat[]{
    \includegraphics[width=0.22\linewidth]{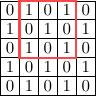}
    \label{fig.KPU_mp_data_order_b}
  }
\caption{The pixel order of the input image when processing two pixels per clock cycle for a $5 \times 5$ input image where the first (a) and second (b) sliding windows indicate how to delay and connect the KPUs to the input signals.} 
\label{fig.KPU_mp_data_order}
\end{figure}

Fig~\ref{fig.KPU_mp_data_order} shows the input image where each pixel contains either a zero or one, indicating if the pixel is transmitted over either $x_{n,0}$ or $x_{n,1}$. 
Considering the time when the last pixel is processed for the first sliding window, which is marked red in Fig.~\ref{fig.KPU_mp_data_order_a}, it can be determined how to delay and connect the two input signals to each multiplier.
The last multiplier with weight $w_8$ is connected to $x_{n,0}$ without any delay.
The other multipliers have a delayed input to make sure that all inputs for this sliding window are processed at the time as the last pixel arrives.
After calculating the first sliding window, this KPU will then calculate the third sliding window, skipping every second sliding window.
Another KPU with a different delay and connectivity pattern is needed to calculate the other sliding windows.
The other KPU is thereby derived by the second sliding window shown in Fig.~\ref{fig.KPU_mp_data_order_b}.
Now, both KPU designs are repeated for each feature map and for each filter.
Or, when performing depthwise convolution with a channel multiplier of one, only once for each feature map.

This approach yields the benefit that, for a stride $s>1$, some KPU designs may not have to be implemented at all. 
For the example shown, with $s=2$, the second KPU would always produce invalid outputs as the calculated sliding windows are all skipped due to the stride setting and can be removed.
A downside of the approach is that the valid output patterns of both KPU designs can be different, leading to additional control circuitry to filter out invalid outputs.
If zero padding is used, the padding select signals are also different for each KPU design. 
But padding select signals and the validity of the KPU outputs can be determined by the position of the sliding window in the feature map, which can be derived by a simple counter in the layer.

\section{Experiments}
All designs from all our conducted experiments were synthesized for an ADM Virtex UltraScale+ (\texttt{xcvu37p-fsvh2892-3-e}) as in \cite{main_ref}. 
We evaluate our approach using MobileNetV1~\cite{mobilenet} and MobileNetV2~\cite{mobilenet2} models, both trained on the ImageNet dataset~\cite{ImageNet}, which enables direct comparison with a large body of existing work.
We first implemented the MobileNetV1 with the same data rate as in \cite{main_ref} for comparison.

Table~\ref{td.compare_to_mainref} shows the resource utilization of both approaches.
%using the current version of the VHDL code generator developed in \cite{main_ref}. %together with the original results from \cite{main_ref}.
It can be seen that the number of LUTs is reduced tremendously by 22\% and BRAM resources by 15\%, while raising the number of registers by 7\% and slightly reducing DSP resources.
These improvements stem from the fact that the proposed approach explores all viable implementations and can therefore select designs that enable larger compressor trees~\cite{8263391} rather than deriving the layer implementation directly from $\rin$.

To further examine the proposed approach, the MobileNetV2 was implemented with different data rates. 
Table~\ref{td.mobnetv2_datarates} shows the resource utilization, maximum frequency, power, latency and throughput of each implementation. 

It can be seen that by processing two pixels (6 features) per clock cycle, a throughput of almost 16k frames per second (FPS) can be reached. 
But the approach also allows for much lower rates down to just a few hundred FPS where, in return, resources are saved.
The last row shows how an implementation where only 3 features are processed over 32 clock cycles results in only 217 FPS where, in return, only 212 DSPs and 30k LUTs are needed.
The table further shows that BRAM utilization is consistently high, independent of the data rate, because all model weights are stored in BRAM.
When the data rate is low, fewer weights have to be loaded every clock cycle, which could allow an offloading of those weights to either DRAM or HBM memory.

Finally, we compare our approach to the state-of-the-art (SOTA). Table~\ref{td.mobnetv2_datarates} also shows SOTA implementations of the same MobileNetV2 CNN model. It can be seen that our multi-pixel accelerator for the MobileNetV2 reaches more than thrice as many FPS as the current SOTA accelerator.

\section{Future Work}
The proposed approach shows promising results for low latency high throughput CNN inference that scales well to moderate sized CNNs.
The experiments highlight the flexibility of the proposed approach across different data rates, with lower data rates resulting in reduced FPGA resource utilization.
However, as most of the BRAM is used to store the weights of the model, the utilization of BRAM is still very high and does not scale well with the data rate.
This can be solved by offloading weights to DRAM for example.

\section{Conclusion}
In this work, an adaptation of a data rate aware CNN hardware accelerator architecture is proposed.
We enhance the architecture design space for multi-pixel processing and demonstrate how an implementation of a MobileNetV2 can reach up to 16,020 frames per second. 
Additionally, we condense the architectural description and reformulate the process of selecting viable implementation parameters, which determine how the layer is reconfigured.

%Future work:
%- In theory and also in practice, there are multiple candidates that all share the lowest possible input data rate


%\section*{Acknowledgment}




\bibliographystyle{IEEEtran}
\bibliography{IEEEabrv,bibliography}
%\begin{thebibliography}{1}


%\end{thebibliography}

\end{document}